# Realizing Magnetoelectric Coupling with Hydroxide as a Knob

J.Y. Ni[1,2], P.S. Wang[1,2], J. L. Lu[1,2], and H. J. Xiang[1,2*]

[1]*Key Laboratory of Computational Physical Sciences (Ministry of Education), State Key Laboratory of Surface Physics, and Department of Physics, Fudan University, Shanghai 200433, P. R. China*

[2]*Collaborative Innovation Center of Advanced Microstructures, Nanjing 210093, P. R. China*

*email: hxiang@fudan.edu.cn

**Abstract**

Materials with a coexistence of magnetic and ferroelectric order (i.e., multiferroics) provide an efficient route for the control of magnetism by electric fields. Unfortunately, a long-sought room temperature multiferroic with strongly coupled ferroelectric and ferromagnetic (or ferrimagnetic) orderings is still lacking. Here, we propose that hydrogen intercalation in antiferromagnetic transition metal oxides is a promising way to realize multiferroics with strong magnetoelectric coupling. Taking brownmillerite $SrCoO_{2.5}$ as an example, we show that hydrogen intercalated $SrCoO_{2.5}$ displays strong ferrimagnetism and large electric polarization in which the hydroxide acts as a new knob to simultaneously control the magnetization and polarization at room temperature. We expect that ion intercalation will become a general way to design magnetoelectric and spintronic functional materials.

Multiferroics [1-8] are attracting increasing interest not only due to the profound underlying physics (i.e., the coexistence and coupling of ferroelectric and magnetic orders), but also the potential applications in novel multifunctional devices. In particular, multiferroics may be used to realize non-volatile, low-power, and high-density memory devices that combine fast electrical writing and magnetic reading. For such purpose, an ideal multiferroic should have large ferroelectric polarization and large spontaneous magnetization at room temperature, and the magnetoelectric (ME) coupling should be strong [9].

Unfortunately, such high-performance multiferroics have not been realized experimentally. In the so-called type-II multiferroics [10-12] (e.g., TbMnO$_3$), the electric polarization caused by the non-centrosymmetric magnetic order is usually too small despite of the intrinsic strong ME coupling. In type-I multiferroics in which the ferroelectricity and magnetism have different sources, the ME coupling is usually weak [9,13]. Interestingly, it was recently proposed [14,15] that the oxygen octahedral rotation can be utilized as a knob to control simultaneously the magnetism (**M**) and polarization (**P**). For instance, tuning the anti-phase octahedral rotation in the hybrid improper ferroelectric material Ca$_3$Mn$_2$O$_7$ can not only affect the polarization arising from the trilinear coupling between polarization and two rotational modes, but also change the canted magnetic moment since the Dzyaloshinskii-Moriya (DM) interaction is correlated to the Mn-O-Mn angle. However, its magnetic critical temperature ($T_N$ =118 K) is low, and the canted magnetic moment is small, and the switch of its ferroelectric polarization has not been achieved experimentally. Hence, it will be interesting from both physics and application points of view to discover new mechanisms for realizing strong ME coupling.

In this Letter, we propose that "hydroxide" (i.e., the O-H covalent bond) can act as a knob to realizing ME coupling. To be more specific, the intercalation of the hydrogen ions can transform an antiferromagnetic (AFM) non-polar material into a multiferroic with a large polarization and a high magnetization, and control the orientation of hydroxide results in a simultaneous change of magnetization and polarization. We demonstrate this general idea by considering the intercalation of hydrogen ions in brownmillerite $SrCoO_{2.5}$.

*General idea*-Our general idea is illustrated in Fig. 1(a). One starts from a parent transition-metal oxide material which displays an AFM magnetic order with a high Néel temperature. Then we insert hydrogen ions into the parent material. The intercalated hydrogen ions will form O-H hydroxides. The hydroxide may deviate from the symmetric middle position, possibly in order to form an O⋯H hydrogen bond with the other neighboring $O^{2-}$ ion. As a result, there could be two possible configurations: hydroxide-downward [left panel of Fig. 1(a)] or hydroxide-upward [right panel of Fig. 1(a)]. Note that if all the transition-metal B ions in the parent material adopt the same valence state (i.e., +n), the $B^{n+}$ ion that is closest to the $H^+$ ion tends to become $B^{(n-1)+}$ to maintain the charge neutrality and minimize the Coulomb interaction. Since the $B^{(n-1)+}$ ion has a different magnetic moment from the $B^{n+}$ ion, these two configurations will have opposite total magnetic moments assuming that the parent material has a robust AFM order (i.e., the flop of the $H^+$ ion does not change the spin directions of all the $B^{n+}$ ions of the material). If these two configurations are insulating, their electric polarizations will be also different due to the different charge ordering and different position of the $H^+$ ion. This suggests that an external electric field can switch between these two configurations. Therefore, hydroxide can serve as a knob to simultaneously manipulate the magnetization and polarization. There are three prerequisites to realize such an idea: (1) identify a parent transition-metal oxide with a robust AFM order, which is easy since most transition-metal oxides display a room temperature AFM order [16] (e.g., the *G*-type order); (2) One can intercalate hydrogen into the parent material; (3) The involved transition-metal ion of the parent transition-metal oxide should have the ability to change from one oxidation state to another, i.e., it presents multiple oxidation states. In the following, we will mainly discuss the ME coupling in hydrogen-intercalated brownmillerite $SrCoO_{2.5}$.

***ME coupling in hydrogen-intercalated $(SrCoO_{2.5})_8H_1$***. We will demonstrate the above idea with brownmillerite $SrCoO_{2.5}$ as the parent material. The brownmillerite structure can be derived from the $ABO_3$ perovskite oxides, by replacing 1/6 of oxygen ions by vacancies. The oxygen vacancies are ordered in such a way so that the tetrahedral $CoO_4$ layer and octahedral $CoO_6$ layer are alternative stacked along the *c*-axis in $SrCoO_{2.5}$. $SrCoO_{2.5}$ displays a robust *G*-type AFM order below 537 K [17-19]. Recently, the ionic liquid gating technique was adopted to insert $H^+$ ions into $SrCoO_{2.5}$ to tune the optical and magnetic properties [19]. All these properties make $SrCoO_{2.5}$ an ideal parent material to realize the electric field control of magnetism with hydroxide as a knob. Note that $SrCoO_{2.5}$ might adopt different phases with varying tilting patterns of the $CoO_4$ tetrahedrons. Among the common two phases (*Pnma* and *Ima2*) [18], our density functional theory (DFT) calculations show that the *Pnma* phase has a lower energy by 54.88 meV/Co. Hereafter, we will mainly focus on the non-polar *Pnma* phase of $SrCoO_{2.5}$. However, our test calculation shows that similar results on the hydroxide knob are obtained in the case of the *Ima2* phase (see Sec. II. of Supplementary material (SM) [20]).

We will first consider the case when one hydrogen is added to the $SrCoO_{2.5}$ system, and then discuss the case of the presence of multiple hydrogens. For simplicity, we denote the structure of the $SrCoO_{2.5}$ unit cell (36 atoms in total) with one additional hydrogen as $(SrCoO_{2.5})_8H_1$. We will mainly report the results obtained with the PBE functional since our tests show that the more accurate and demanding optB88_vdW functional [21] gives similar results (see Sec.III.1 of SM [20]). We determine the ground state structure of $(SrCoO_{2.5})_8H_1$ by considering all the possible absorption oxygen sites. It turns out that the hydrogen ion will form a covalent O-H bond (i.e., hydroxide) with the bridge oxygen ion between the $CoO_4$ tetrahedron and the $CoO_6$ octahedron, in agreement with the previous result [19]. The formation energy calculations show that the hydrogen intercalation in $SrCoO_{2.5}$ is thermodynamically favorable [22], in agreement with the experimental observation [19]. We find that there are several locally stable states when the hydrogen ion is absorbed to the bridge oxygen since the hydroxide can rotate with respect to the oxygen to form a hydrogen bond with the other oxygen ions[20]. In particular, we show in Fig. 1(b) the two lowest energy states in which the hydroxide is orientated toward the lower $CoO_4$ tetrahedron (referred to as tet1) or toward the upper $CoO_6$ octahedron (referred to as oct1), respectively. The tet1 state has a slightly lower energy than the oct1 state by 20 meV/H. For both states, the bond lengths of the O−H hydroxide and the O···H hydrogen bond are about 1 Å and 2 Å, respectively. Our first-principles molecule dynamics (MD) simulations indicate that tet1 and oct1 states are thermally stable at room temperature[20]. Our energy barrier calculations suggest that it is relatively easier to adjust the orientation of the hydroxide than to change the absorbed oxygen site of the hydrogen[20]. This is understandable since breaking the O···H hydrogen bond is much easier than breaking the covalent O-H bond.

In the parent material $SrCoO_{2.5}$, all the $Co^{3+}$ ions adopt the high-spin configuration (i.e., the local magnetic moment is close to 4 $\mu_B$). Our calculations show that the magnetic ground state of $SrCoO_{2.5}$ is the *G*-type magnetic ordering, in agreement with the experimental results [17-19]. For tet1 (oct1) configuration, the tetrahedral (octahedral) Co ion close to the hydroxide becomes the $Co^{2+}$ ion. We find that the $Co^{2+}$ ion also adopts a high-spin configuration with the local magnetic moment close to 3 $\mu_B$. The charge ordering pattern can be also seen from the density of state (DOS) analysis (see Fig. 2(b)). We compute the spin exchange interactions (Table S1-2) with the four-state mapping method [23,24] to find that the $Co^{2+}$ spin couples antiferromagnetically with the neighboring $Co^{3+}$ spins. In fact, both tet1 and oct1 configurations adopt the *G*-type magnetic ordering [20]. Interestingly, the *G*-type magnetic ordering for both tet1 and oct1 configurations is ferrimagnetic instead of the usual AFM (e.g., $SrCoO_{2.5}$ [18]): Our DFT+*U* collinear spin-polarized calculation shows that the net magnetizations of the tet1 and oct1 configurations are 1 $\mu_B$ and -1 $\mu_B$ per unit cell, respectively. This is because the $Co^{2+}$ ion has a smaller local moment than the $Co^{3+}$ ion, and the neighboring Co spins are antiparallel to each other. By performing the Monte Carlo (MC) simulations, we estimate that the ferrimagnetic Curie temperature of oct1 and tet1 configurations are all above 500 K [20].

We further consider the spin-orbit coupling (SOC) effect to examine the magnetic anisotropy to find that the magnetic easy-axes are along *c*-axis for purely $SrCoO_{2.5}$ and oct1 configuration of $(SrCoO_{2.5})_8H_1$, while the magnetic easy-axis of the tet1 configuration is along *a*-axis. To clearly understand how the orientation of the hydroxide affects the global magnetic anisotropy, we calculate single ion anisotropy (SIA) parameters of cobalt ions in $SrCoO_{2.5}$ and two configurations of $(SrCoO_{2.5})_8H_1$, as shown in the Table S3-5. For $SrCoO_{2.5}$, the SIA parameters show that the tetrahedral $Co^{3+}$ ions have strong magnetic easy-axis behavior along *c*-axis, while the spins of the octahedral $Co^{3+}$ ions prefer to be in the *ab*-plane. Thus, there is a competition in the magnetic anisotropy between the two kinds of $Co^{3+}$. The reason why the magnetic easy-axis is along the *c*-axis for $SrCoO_{2.5}$ is that the tetrahedral $Co^{3+}$ ions have stronger magnetic anisotropy. For the magnetic anisotropy property in $(SrCoO_{2.5})_8H_1$, we find that the absorbed hydrogen ion mainly changes the SIA parameters of the two Co ions (Co1 and Co2 in Fig. S10) that bond with the oxygen ion of the hydroxide. For the oct1 configuration of $(SrCoO_{2.5})_8H_1$, the easy axis is still along the *c*-axis despite of the fact the magnetic easy-plane anisotropy of the octahedral Co2 ($Co^{2+}$) ion becomes stronger. However, in the tet1 configuration, the magnetic anisotropy of the $Co^{2+}$ ion (i.e., Co1 in Fig. S10) near the hydroxide changes: the easy axis is now along *a*-axis instead of *c*-axis. As a result, the magnetic easy-axis of tet1 configuration of $(SrCoO_{2.5})_8H_1$ is now along *a*-axis instead of *c*-axis. Therefore, the change of magnetic anisotropy of $(SrCoO_{2.5})_8H_1$ can be seen as a consequence of the local effect of the hydrogen intercalation.

Our calculations show that both tet1 and oct1 configurations have a relatively large band gaps (0.95 eV and 0.74 eV, respectively). Since these two configurations differ in the orientation of the hydroxide and charge order pattern, there will be a non-zero electric polarization difference between the two configurations. With the Berry phase method [25], the difference in the out-of-plane electric polarization between tet1 and oct1 configurations is computed to be 18.79 $\mu C/cm^2$ along the [001] direction. This polarization value is close to that of proper ferroelectrics (e.g., $BaTiO_3$) [26], which is much larger than that of type-II multiferroics [10]. Note that the direction of the polarization is opposite to the orientation of the hydroxide: If we only consider the polarization contributed from the hydroxide, the difference in the out-of-plane electric polarization between tet1 and oct1 configurations should be along the [00-1] direction since in the tet1 (oct1) case, the hydroxide is orientated downward (upward). This is because the polarization difference is in fact mainly contributed by the charge ordering instead of the hydroxide position due to the large distance between the $Co^{2+}$ and $Co^{3+}$ ion along the $c$-axis. Our simple estimation (see Fig. 2(a)) shows that the electric polarization difference contributed from the rotation of the hydroxide is about -1 e·Å (-3.41 $\mu C/cm^2$), while the contribution from the charge ordering is 4.5 e·Å (15.35 $\mu C/cm^2$).

Our above results indicate that $(SrCoO_{2.5})_8H_1$ has two locally stable configurations with different magnetic properties and polarization associated with the different orientation of the hydroxide. We expect that an external electric field along the [001] direction can switch between the oct1 configuration and tet1 configuration due to the polarization difference. Due to the ME coupling mediated by the orientation of the hydroxide, the electric field can also switch the easy-axis of the ferrimagnetic state. As these two configurations are not symmetrically equivalent, $(SrCoO_{2.5})_8H_1$ can be regarded as an asymmetric multiferroic [27] with a large magnetization and large electric polarization at room temperature. Note that here the ME coupling in $(SrCoO_{2.5})_8H_1$ with a fixed hydrogen amount is different from the phase transformation between AFM hydrogen-free $SrCoO_{2.5}$ and weakly ferromagnetic $(SrCoO_{2.5})_8H_8$ induced by the electric-field [19].

Next, let us consider the case of inserting two hydrogens into $SrCoO_{2.5}$. In the case of the lowest energy absorption pattern, there are four stable configurations which are named as TT, TO, OT and OO configurations, respectively (detailed results of $(SrCoO_{2.5})_8H_2$ are given in the Sec. IV. of SM). All these four configurations adopt $G$-type magnetic order as ground state. However, they adopt different charge ordering ground states: TO and OT configurations display the $G$-type charge ordering, while TT and OO configurations display the $C$-type charge ordering. Considering the charge order and spin order, the net magnetizations for TT, TO, OT, and OO configurations are 0 $\mu_B$, -2 $\mu_B$, 2 $\mu_B$, and 0 $\mu_B$ per unit cell, respectively. The magnetic easy axis is found to be along the $a$-axis for all four configurations. As shown in Fig. 3, four configurations of $(SrCoO_{2.5})_8H_2$ are very close in energy and four configurations have different magnetic and ferroelectric properties, suggested that $(SrCoO_{2.5})_8H_2$ can be used as multiple-states memory, where the transformation between these states can be achieved with an electric field. Because of the entropy effect, the absorbed hydrogen ions may be disordered. To examine the effect of the disorder, we also consider the other two cases where the two hydroxides in $(SrCoO_{2.5})_8H_2$ are far from each other [20]. In these cases, we also find that hydroxide can act as a knob to simultaneously control the polarization and magnetization.

*Discussion*-The multiferroic mechanism proposed in this work differs from the usual known mechanisms in several aspects: (1) The local electric dipole in hydrogen intercalated AFM oxides is due to the orientation of the hydroxide and the associated charge order, resulting in a strong ME coupling. Although hydrogen-bond ordering was found to induce ferroelectricity in multiferroic systems such as $[(CH_3)_2NH_2]Mn(HCOO)_3$ [28], the magnetism hardly couples with the ferroelectricity. (2) Although charge order was suggested to induce ferroelectricity in some multiferroics [29-31], the ferroelectricity is hardly switchable by an external electric field since the conductivity in these systems appears to be too high due to a small band gap. In hydrogen intercalated AFM oxides, the band gap could be rather large since the $H^+$ ion provides an additional attractive potential for the electron of the transition-metal ion close to the $H^+$ ion. For example, the band gap of $(SrCoO_{2.5})_8H_2$ is computed to be 1.2 eV. (3) A collective ordering of the hydroxide-related local electric dipoles in hydrogen intercalated AFM oxides is not necessary for realizing the coupling between the ferrimagnetism and electric dipole since an isolated hydroxide in an AFM oxide is itself stable at room temperature. This may suggest a new route to high-density storage with a single isolated hydroxide representing one bit. In contrast, the usual multiferroic displays a conventional ferroelectricity with a collective ordering of the local dipoles. (4) Besides $SrCoO_{2.5}$, we also find that hydroxide can act as a knob to simultaneously control the magnetization and polarization in other hydrogen intercalated compounds (such as $CaFeO_{2.5}$ and perovskite $LaFeO_3$), indicating that proposed ME coupling mechanism is general [32].

To summarize, we introduce a new knob - "hydroxide" to simultaneously control magnetization and electric polarization in AFM transition metal oxides. As we demonstrated in the cases of hydrogen intercalated $SrCoO_{2.5}$, this idea leads to a new way to realize room temperature multiferroics in which the strong electric polarization and high magnetization are tightly coupled with each other. Our study suggests that ion intercalation is a promising way to rationally design new functional magnetoelectric and spintronic materials.

**ACKNOWLEDGMENTS**

This work is supported by the Special Funds for Major State Basic Research (Grant No. 2015CB921700), the Qing Nian Ba Jian Program, and the Fok Ying Tung Education Foundation.

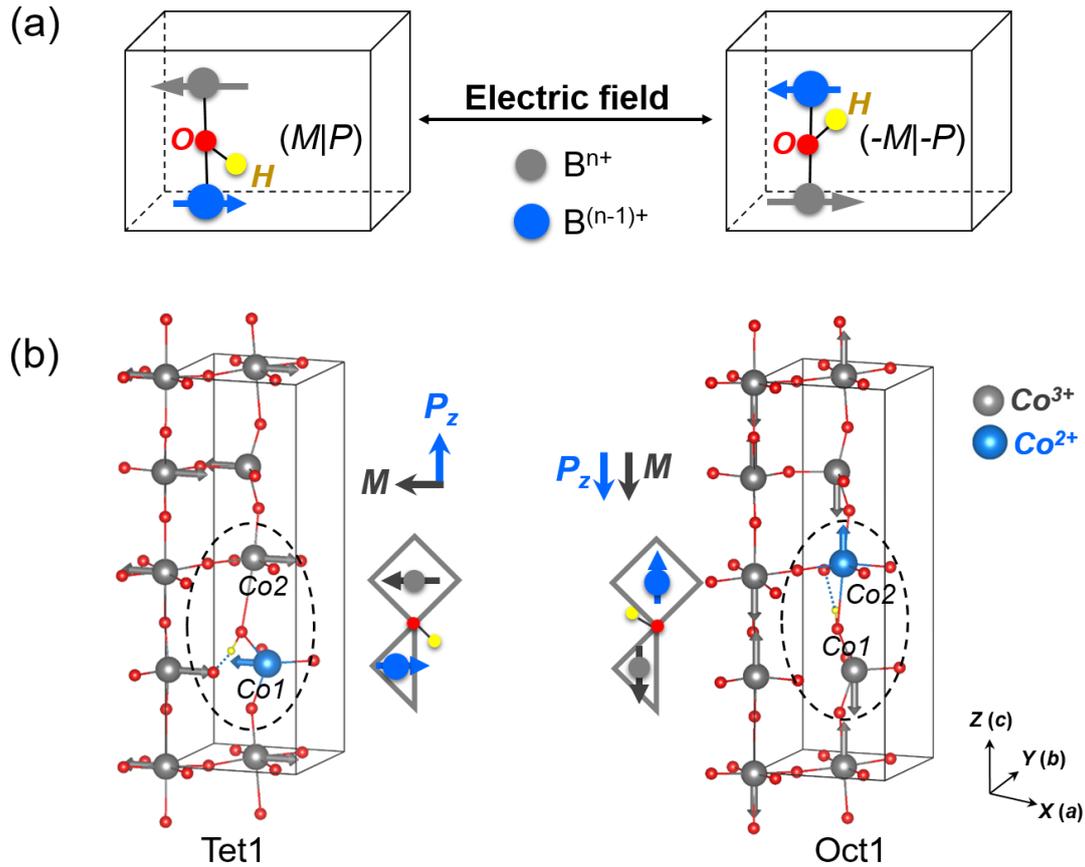

**FIG. 1.** (a) Illustration of the general idea of using hydroxide as a knob to simultaneously control the magnetization and polarization. Longer gray arrow and shorter blue arrow represent local magnetic moment of magnetic ions ($B^{n+}$ and $B^{(n-1)+}$), respectively. For clarity, only the two B-ions near the hydroxide are shown. (b) ME coupling in $(SrCoO_{2.5})_8H_1$. Depending on the orientation of the hydroxide, two different configurations (left: tet1 configuration, right: oct1 configuration) have different net magnetization (M) and polarization (P). The local magnetic moments for each Co ion are displayed. The rhombus and triangle represent the oxygen octahedron and tetrahedron, respectively. Hydroxide and hydrogen bond are represented by solid and dotted line, respectively.

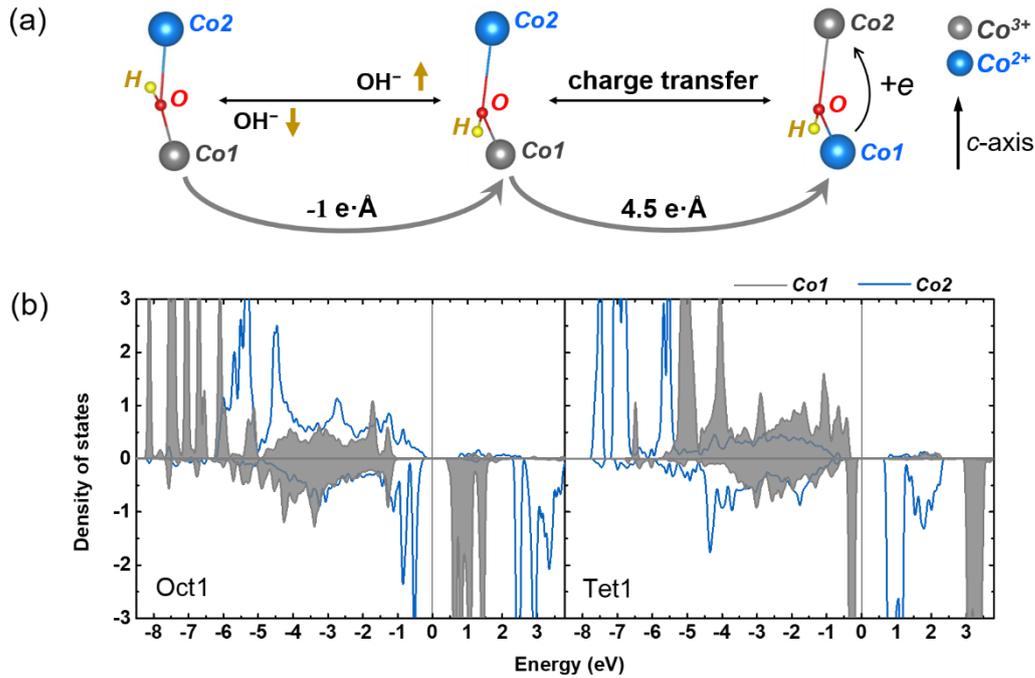

FIG. 2. Origin of the electric polarization difference between tet1 and oct1 configurations. (a) Simple estimation of the polarization change due to the hydroxide rotation and charge transfer associated with the charge ordering. Tetrahedral Co1 and octahedral Co2 are labeled in Fig. 1(b). (b) The partial DOS plot of the tetrahedral site Co1 (gray line) and the octahedral site Co2 (blue line) in the oct1 configuration (left) and tet1 configuration (right), evidencing the charge transfer associated with the charge ordering. In the DOS calculations, the $G$-type magnetic ordering is adopted. The positive and negative values represent majority component and spin minority component, respectively.

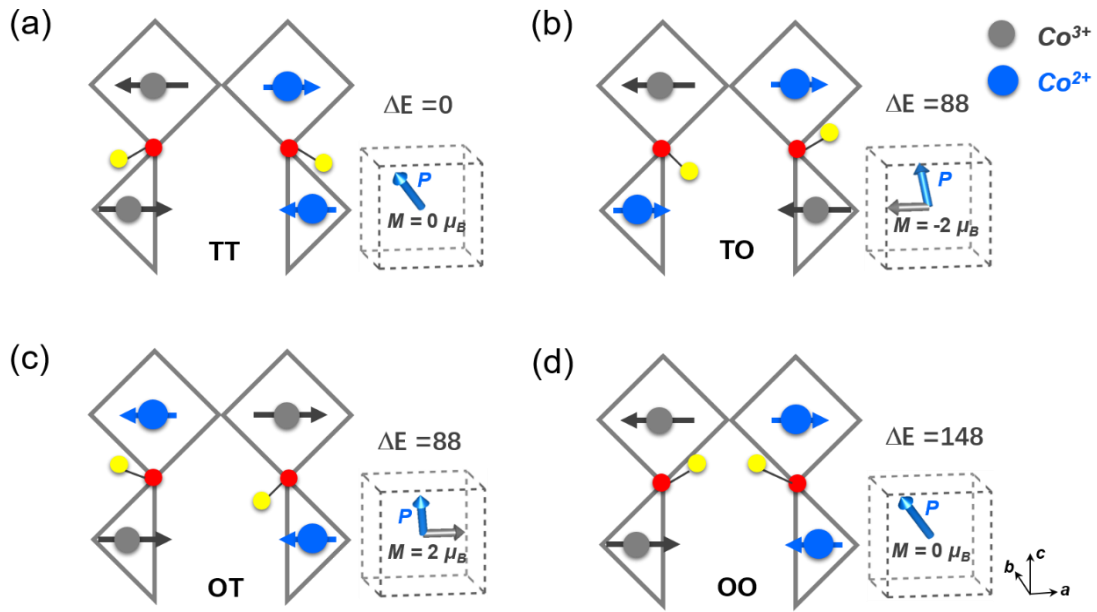

**FIG. 3.** Schematics of four configurations in $(SrCoO_{2.5})_8H_2$. (a) TT configuration: both hydroxides point toward oxygen tetrahedron, (b)-(c) TO and OT configuration: one hydroxide point toward the octahedron, while the other toward the tetrahedron, (d) OO configuration: both hydroxides point toward oxygen octahedron. All four configurations adopt the G-type magnetic structure with the easy-axis along the *a*-axis as the magnetic ground state. Arrows denote the spin directions of the Co ions. Relative energies (meV/2H), net magnetization and direction of total electric polarization are given for different configurations.